\documentclass[12pt]{article}
\usepackage{amsfonts}
\usepackage{mathrsfs}
\usepackage{amsthm}
\usepackage{multirow}
\usepackage{bbding}
\usepackage{amssymb}
\usepackage{amsmath}
\usepackage{graphicx,color}
\usepackage{rotating}

\usepackage{bm}
\usepackage{epstopdf}
\newcommand{\bea}{\begin{eqnarray}}
\newcommand{\eea}{\end{eqnarray}}
\newcommand{\Bea}{\begin{eqnarray*}}
\newcommand{\Eea}{\end{eqnarray*}}
\newcommand{\ba}{\begin{array}}
\newcommand{\ea}{\end{array}}
\newcommand{\bt}{\begin{tabular}}
\newcommand{\et}{\end{tabular}}
\newcommand{\btb}{\begin{table}}
\newcommand{\etb}{\end{table}}
\newcommand{\bc}{\begin{center}}
\newcommand{\ec}{\end{center}}
\newcommand{\beq}{\begin{equation}}
\newcommand{\eeq}{\end{equation}}

\usepackage{float}

\makeatletter

\newcommand{\Rmnum}[1]{\expandafter\@slowromancap\romannumeral #1@}
\makeatother

\begin{document}

\title{Unified Rules of Renewable Weighted Sums for Various Online Updating Estimations
}
\author{Lu Lin$^1$, Weiyu Li$^1$\footnote{The corresponding
author. Email: liweiyu@sdu.edu.cn. The research was
supported by NNSF projects (11971265) of China.} \ and Jun Lu$^2$
\\
\small $^1$Zhongtai Securities Institute for Financial Studies, Shandong University, Jinan, China\\
\small $^2$School of Statistics and Mathematics, Zhejiang Gongshang University, Hangzhou, China
 }
\date{}
\maketitle

\vspace{-0.7cm}

\begin{abstract} \baselineskip=17pt

This paper establishes unified frameworks of renewable weighted sums (RWS) for various online updating estimations in the models with streaming data sets. The newly defined RWS lays the foundation of
online updating likelihood, online updating loss function, online updating estimating
equation and so on. The idea of RWS is intuitive and heuristic, and the algorithm is computationally simple. This paper chooses nonparametric model as an exemplary setting. The RWS applies to various types
of nonparametric estimators, which include but are not limited to nonparametric likelihood, quasi-likelihood and least squares. Furthermore, the method and the theory can be extended into the models with both parameter and  nonparametric function. The estimation consistency and
asymptotic normality of the proposed renewable estimator are established, and the oracle property is obtained. Moreover, these properties are always satisfied, without any constraint on the number of data batches, which means that the new method is adaptive to the situation where
streaming data sets arrive perpetually. The behavior of the method is further illustrated by various numerical
examples from simulation experiments and real data analysis.

{\it Key words: Streaming data set; Online learning; Online updating likelihood; Online updating estimating equation; Nonparametric estimation.}

\end{abstract}

\baselineskip=20pt

\setcounter{equation}{0}
\section{Introduction}

Streaming data sets are a special data type in the emerging field of ``big data". In such a data environment, data arrive in streams and chunks, and a main issue is how to address statistics in an online updating framework, without storage requirement for previous raw data. Up to now, various statistical and computing methodologies that enable us to sequentially
update certain statistics have been proposed in the existing literature. The examples include recursive operation, stochastic gradient descent algorithm, online second-order method, online Bayesian inference and so on (the references will be given later). Even so, however, there are still the following challenging issues:
\begin{itemize}
\item[1)] Most existing methodologies are developed from case to case, or are adaptive merely to certain settings. It is then desired to develop relatively unified notion and strategy, based on which one can construct online updating statistics for general models with streaming data sets. This idea is similar to the classical rules for statistical inference, such as likelihood, loss function and estimating equation and so on. Thus, the general rules we try to develop in the area of streaming data sets are, for example, the online updating likelihood, online updating loss function and online updating estimating equation.
\item[2)] Most existing methodologies need strong constraints on the total number
of streaming data sets to achieve the statistical consistency and oracle property; more specifically, a commonly used constraint has the form of $k=O(n^c)$ for some constant $0<c<1$, where $n$ is the total number of data and $k$ is the total number of data batches. So, it is also desired that the unified strategies are free of the constraint and then are adaptive to the
situation where streaming data sets arrive fast and perpetually.
\end{itemize}

To the best of our knowledge, the issue about the unified rules aforementioned in 1) has rarely been investigated, and the issue about how to remove the constraint $k=O(n^c)$ as in 2) is still a big challenge.
In this paper, we focus on the two issues and introduce the following unified rules for statistical estimation under general models with streaming data sets.
\begin{itemize}
\item {\it The framework of online updating loss function.} The proposed rule in this paper has the form of
\begin{equation}\label{(intro-0)}\widehat\theta_k=\min_{\theta}\sum_{j=1}^{k-1}\sum_{i\in {\bm i}_j}
    J(\widehat\theta_j;Z_i)D(\theta,\widehat\theta_{k-1})+
    L(\theta;{\bm d}_k),\end{equation}
where $\theta$ is an unknown parameter or nonparametric function to estimate, $\widehat\theta_j$ is the estimator of $\theta$ at the $j$-th updating step, $D(\theta,\widehat\theta_{k-1})$ stands for a distance between $\theta$ and $\widehat\theta_{k-1}$, $J(u;\cdot)$ and $L(u;\cdot)$ are known weight function and loss function respectively, and ${\bm d}_j=\{Z_i:i\in{\bm i}_j\}$ are sequential data sets with the index sets ${\bm i}_j,j=1,\cdots,k$. The
above is of online updating form because it only involves the
current data ${\bm d}_k$, the previous estimator $\widehat\theta_{k-1}$ together with the accumulative  quantity $\sum_{j=1}^{k-1}\sum_{i\in {\bm i}_j} J(\widehat\theta_j;Z_i)$.
When the distance is chosen as $l_2$-norm $D(\theta,\widehat\theta_{k-1})=\|\theta
    -\widehat\theta_{k-1}\|^2$, the rule has the following form:
\begin{equation}\label{(intro-1)}\widehat\theta_k=\min_{\theta}\sum_{j=1}^{k-1}\sum_{i\in {\bm i}_j}
    J(\widehat\theta_j;Z_i)\|\theta
    -\widehat\theta_{k-1}\|^2+
    L(\theta;{\bm d}_k).\end{equation}
\item {\it The framework of online updating estimating equation.} The proposed rule in this paper has the form that the online updating estimator $\widehat\theta_k$ is the solution to the following equation:
   \begin{equation}\label{(intro-0-1)}\sum_{j=1}^{k-1}\sum_{i\in {\bm i}_j}
    J(\widehat\theta_j;Z_i)U_0(\theta-\widehat\theta_{k-1})-
    U(\theta;{\bm d}_k)=0\end{equation} for $\theta$, where function $U_0(u)$ satisfies $U_0(0)=0$, and $U(\theta;\cdot)$ is an unbiased estimating function of $\theta$.
    When the function $U_0(u)$ is chosen as $U_0(u)=u$, the estimating equation has the following form:
    \begin{equation}\label{(intro-2)}\sum_{j=1}^{k-1}\sum_{i\in {\bm i}_j}
    J(\widehat\theta_j;Z_i)(\theta-\widehat\theta_{k-1})-
    U(\theta;{\bm d}_k)=0\end{equation} for $\theta$.
\end{itemize}
In this paper, we mainly focus on the rules (\ref{(intro-1)}) and (\ref{(intro-2)}) for simplicity.
The two rules in (\ref{(intro-1)}) and (\ref{(intro-2)}) are equivalent under some regularity conditions.
We call the rules in (\ref{(intro-1)}) and (\ref{(intro-2)}) as ``renewable weighted sum", denoted by RWS for short, because of the renewability:
the $k$-th updating procedure only uses the current loss function $L(\theta;{\bm d}_k)$ or the current estimating function $U(\theta;{\bm d}_k)$, the last estimator $\widehat\theta_{k-1}$ and the accumulative weight $\sum_{j=1}^{k-1}\sum_{i\in {\bm i}_j}J(\widehat\theta_j;Z_i)$ of the previous weights $J(\widehat\theta_j;Z_i)$ for $j=1,\cdots,k-1$, instead of the previous raw data sets ${\bm d}_j$ for $j=1,\cdots,k-1$.
It will be shown in Section 3 that the RWS is theoretically reasonable in the following perspective:
\begin{itemize}
\item
 The RWS is an online updating likelihood or online updating score under some cases (for the details see Remark 1 given in Section 3). Consequently, the classical theoretical properties, such as estimation efficiency and oracle property, can be successfully achieved.  \end{itemize}
The rules of the RWS are intuitive and heuristic, more specifically, the online updating estimator $\widehat \theta_k$ should be close to the previous estimator $\widehat \theta_{k-1}$ and should approximately minimize the current loss function $L(\theta;{\bm d}_k)$ in (\ref{(intro-1)}) or should approximately satisfy the current estimating equation $U(\theta;{\bm d}_k)=0$ in (\ref{(intro-2)}).

In this paper, we choose  nonparametric model as an exemplary setting, and focus on its estimation method, algorithm and theoretical properties. Our nonparametric method has the following salient features:
\begin{itemize}
\item[a)] {\it Unified framework.} The RWS applies to various types
of nonparametric estimators, which include but are not limited to nonparametric likelihood, quasi-likelihood and least squares. Moreover, the estimation method and theoretical conclusion can easily be extended into both parametric and semiparametric models. Actually, the proposed RWS is  a general rule of nonparametric online updating likelihood, nonparametric online updating loss function and nonparametric online updating estimating equation for general statistical estimations.
\item[b)] {\it Proporties of efficiency, oracle and adaptability.} The resulting online updating estimator has estimation efficiency, asymptotic normality and oracle property. Moreover, these properties always hold for any number of data batches, without the constraint of $k=O(n^c)$. Therefore, the online updating estimation is adaptive to the situation where
streaming data sets arrive fast and perpetually.
\end{itemize}

For better understanding our methodological development, in the following, we briefly summarize the related works on the statistical and computing methodologies in the area of streaming data sets.

The modern developments of science and technology have enabled the massive data sets arising in various fields. The major challenge in analyzing this kind of data is that data storage and analysis by standard computers are hardly feasible. Up to now, there haven been three main strategies to deal with the problem: sub-sampling (see, e.g., Liang et al., 2013; Kleiner et al., 2014;
Maclaurin and Adams, 2014; Ma, Mahoney and Yu, 2015), divide and conquer (see, e.g., Lin
and Xi, 2011; Neiswanger,Wang and Xing, 2013; Scott et al., 2013; Chen and Xie, 2014; Song and Liang,
2014; Pillonetto, et al., 2019), and the online updating (see, i.e., Schifano et al., 2016; Wang et al, 2018; Xue et al, 2019; Luo and Song, 2020). The online updating approach, however, is basically distinct from the other two because the data sets arrive in streams and
chunks, and the statistical method should be of an online updating framework, without storage requirement
for historical data.

Up to now, there have been several online updating approaches to analyzing streaming data sets, for example,  recursive operation, stochastic gradient descent algorithm, online second-order method and online Bayesian inference. In some simple cases, such as sample mean, least squares
estimator in linear regression and N-W estimator in nonparametric regression, the previous statistics can be updated directly with new data set by recursive operation (see, e.g., Schifano, et al., 2016). This simple
strategy has been widely used in the existing literature (see, e.g., Bucak
and Gunsel, 2009; Nion and Sidiropoulos, 2009). In most situations, however, the statistics are not a linear function of data, and moreover, often have no
closed form expression. In these complicated situations, the online updating solution only can be obtained numerically by iterative
algorithms, such as the Newton-Raphson algorithm. Thus, stochastic gradient descent algorithm and its improved versions can be used to update the statistics with sequentially arriving data (see, e.g., Robbins and Monro, 1951; Bordes et al., 2009; Duchi et al.,
2011; Toulis et al., 2015). Another widely used updating scheme is online second-order methods such as the natural gradient algorithm and the online Newton step (see, e.g., Amari
et al., 2000;  Hazan et al., 2007; Vaits et al., 2015;  Hao et al., 2016). For an extended version, online quasi-Newton method, see, e.g., Nocedal and Wright (1999), Liu and Nocedal (1989), Schraudolph et al. (2007) and Bordes et al. (2009).

To achieve the oracle property and guarantee the estimation consistency, however, most existing methodologies need strong constraints on the total number
of streaming data sets, such as $k=O(n^c)$, where $0<c<1$ is a constant, $n$ is the number of total data and $k$ is the total number of the data batches. Thus, these methodologies are not adaptive to the
situation where streaming data sets arrive fast and perpetually. Recently, Luo and Song (2020) proposed an incremental updating algorithm by Taylor series
expansion of score function in generalized linear models, the resulting approximate score function is similar to (\ref{(intro-1)}) and (\ref{(intro-2)}), and the constraint can be relaxed. On the other hand, most existing methodologies in the area of streaming data sets are developed from case to case, implying they are adaptive merely to certain circumstances.
The above observations indicate that it is desired to develop unified
notion and strategy such that one can construct the online updating statistics for general models
with streaming datasets, and can remove the constraint condition. These are the main targets we will achieve in this paper.

The remainder of this paper is organized in the following way. In Section 2, some motivating examples are discussed to initiate the methodological development. In Section 3, a unified framework of RWS is introduced, correspondingly, the online updating estimating equation and online updating loss function are proposed, and the algorithms are suggested. The theoretical properties of the online updating estimator are investigated in Section 4. The main simulation studies and real data analysis are provided in Section 5 to
illustrate the new method. Proofs of the theorems and some further simulation studies are relegated to Supplemental Materials.

\setcounter{equation}{0}
\section{Motivating examples}

To proceed with the methodological development,
we first look at some motivating examples.
Consider the following nonparametric regression:
\begin{equation}\label{(regression)}Y_i=r(X_i)+\varepsilon_i,\end{equation} in which the data sets are available
sequentially as ${\bm d}_j=\{(Y_i,X_i):i\in{\bm i}_j\}$
with index sets ${\bm i}_j$ and $j=1,2,\cdots $. Denote ${\cal D}_k=\bigcup_{j=1}^k{\bm d}_j$ and ${\cal I}_k=\bigcup_{j=1}^k{\bm i}_j$. In this paper, we suppose that  $(Y_i,X_i)^T$ for all $i$ are independent and identically distributed observations of $(Y,X)^T$, and $Y$ and $X$ are scalar response and covariate respectively, only for the simplicity of representation. By Nadaraya (1964) and Watson (1964), the N-W estimator $r(x)$ computed on ${\cal D}_k$ is defined by
\begin{equation}\label{(regression-estimator)}\widehat r_k(x)=\frac{\sum_{i\in {\cal I}_k}Y_iK_{h}(X_i-x)}{\sum_{i\in {\cal I}_k}K_{h}(X_i-x)},\end{equation} where $K_{h}(\cdot)=\frac 1{h}K\left(\frac{\cdot}{h}\right)$ with $K(\cdot)$ and $h$ be kernel function and bandwidth, respectively. It can be verified that the estimator $\widehat r_k(x)$ satisfies the following equation:
\begin{equation}\label{(updating)}\sum_{i\in {\cal I}_{k-1}}K_{h}(X_i-x)(\widehat r_k(x)-\widehat r_{k-1}(x))-\sum_{i\in {\bm i}_k}(Y_i-\widehat r_k(x))K_{h}(X_i-x)=0,\end{equation} where the initial estimator is $\widehat r_0(x)\equiv0$ by convention. Then, the estimator $\widehat r_k(x)$ could be thought of as the online updating version from the prior estimator $\widehat r_{k-1}(x)$ in the sense of estimating equation. In other words, $\widehat r_k(x)$ is the solution to the incremental equation:
\begin{equation}\label{(updating-1)}\sum_{i\in {\cal I}_{k-1}}K_{h}(X_i-x)( r(x)-\widehat r_{k-1}(x))-\sum_{i\in {\bm i}_k}(Y_i-  r(x))K_{h}(X_i-x)=0 \end{equation} for $r(x)$, or equivalently, it is the solution to the incremental optimization problem:
\begin{equation}\label{(updating-2)}\min_{r(x)}\sum_{i\in {\cal I}_{k-1}}K_{h}(X_i-x)( r(x)-\widehat r_{k-1}(x))^2+\sum_{i\in {\bm i}_k}(Y_i-  r(x))^2K_{h}(X_i-x).\end{equation}
Actually (\ref{(updating-1)}) and (\ref{(updating-2)}) are newly defined online updating estimating equation and online updating loss function, respectively. They only involve the
current data set ${\bm d}_k$ and the previous estimator $\widehat r_{k-1}(x)$ together with the accumulative  quantity $\sum_{i\in {\cal I}_{k-1}}K_{h}(X_i-x)$.
The incremental updating procedures in (\ref{(updating-1)}) and (\ref{(updating-2)}) imply that the online updating estimator $\widehat r_k(x)$ should be close to the previous estimator $\widehat r_{k-1}(x)$ and should approximately satisfy the current estimating equation $\sum_{i\in {\bm i}_k}(Y_i- r(x))K_{h}(X_i-x)=0$ in (\ref{(updating-1)}) or should approximately minimize the current loss function $\sum_{i\in {\bm i}_k}(Y_i-  r(x))^2K_{h}(X_i-x)$ in (\ref{(updating-2)}).

Moreover, the above incremental updating procedures can be extended into general cases, for example, least squares estimation in linear models, spline estimation and local polynomial estimation in nonparametric models and so on. The details are omitted here.

This idea may date back to the updating weighted sum of Lin and Zhang (2002) as
\begin{equation}\label{(updating-3)}\widehat\theta_2({\bm d}_1,{\bm d}_2)=\arg\max_{\theta}\{l(\theta;{\bm d}_2)-(\widehat\theta_1({\bm d}_1)-\theta)^T
A(\widehat\theta_1({\bm d}_1)-\theta)\},\end{equation}
where $\theta$ is an unknown parameter vector, $l(\theta;{\bm d}_2)$ is the likelihood of $\theta$ computed on the current dataset ${\bm d}_2$, $\widehat\theta_1({\bm d}_1)$ is a previous estimator of $\theta$ computed on the historical dataset ${\bm d}_1$, and $A$ is a weight matrix proportional to the inverse of the asymptotic covariance of $\widehat\theta_1({\bm d}_1)$.
This strategy aggregates the weighted least squares centralized at the previous estimator $\widehat\theta_1({\bm d}_1)$ and the current likelihood $l(\theta;{\bm d}_2)$ respectively from two samples ${\bm d}_1$ and ${\bm d}_2$ to construct an updating likelihood and then get an updating estimator $\widehat\theta_2({\bm d}_1,{\bm d}_2)$. As stated in Introduction, Luo and Song (2020) proposed a similar incremental updating algorithm for parameter estimation by Taylor series
expansion of score function in generalized linear models.

The updating procedures in (\ref{(updating-1)}), (\ref{(updating-2)}) and (\ref{(updating-3)}) can be classified into the unified framework of renewable weighted sum (RWS, for short) as those in (\ref{(intro-1)}) and (\ref{(intro-2)}). This observation motivates us to develop unified rules for constructing online updating estimations under general models with streaming data sets.

\setcounter{equation}{0}
\section{Renewable weighted sums and online updating estimations}

\subsection{Methods in nonparametric models}
We now consider a general case where the true nonparametric function, denoted by $\alpha^0(x)$, is defined as the solution to the conditional estimating equation as
\begin{equation}\label{(estimating equation)}E(U(\alpha(X);Y,X)|X=x)=0, \ x\in [0,1],\end{equation} for nonparametric function $\alpha(x)$. Here $Y$ and $X$ are supposed to be continuous random variables, for simplicity. The function $U(\alpha(x);y,x)$ in the above equation contains the unbiased estimating functions from likelihood, quasi-likelihood and least squares as its special cases, and function $\alpha(x)$ may be the nonparametric regression function $r(x)$ or nonparametric variance function $\sigma^2(x)$ in regression model (\ref{(regression)}).

In the following, we only use kernel estimation as an exemplary method. The method can be extended into the other nonparametric methods such as spline estimation and finite-dimensional approximations (Pillonetto, et al., 2019).
When data sets ${\bm d}_j=\{(Y_i,X_i):i\in{\bm i}_j\},j=1,\cdots,k$, are sequentially observed, motivated by the RWS in (\ref{(updating-1)}), (\ref{(updating-2)}) and (\ref{(updating-3)}), we suggest the incremental updating estimating equation as
\begin{eqnarray}\label{(estimating equation-1)}\nonumber&&\sum_{j=1}^{k-1}\sum_{i\in {\bm i}_j}J(\widehat \alpha_j(x);Y_i,X_i) K_{h_j}(X_i-x)( \alpha(x)-\widehat \alpha_{k-1}(x))\\ && -
\sum_{i\in {\bm i}_k}U(\alpha(x);Y_i,X_i)K_{h_k}(X_i-x)=0 \end{eqnarray} for $\alpha(x)$, where $\widehat \alpha_j(x)$ is kernel estimator with bandwidth $h_j$ obtained at the $j$-th step updating, the initial estimator is chosen as $\widehat \alpha_0(x)\equiv0$ by convention, and $J(\alpha;\cdot,\cdot)$ is a positive weight function. The above is of an online updating form because it only involves the
current data set ${\bm d}_k$ and the previous estimator $\widehat \alpha_{k-1}(x)$ together with the accumulative  quantity $\sum_{j=1}^{k-1}\sum_{i\in {\bm i}_j}J(\widehat \alpha_j(x);Y_i,X_i) K_{h_j}(X_i-x)$.
It can be seen from the proof of Theorem 2 given in Supplementary Materials that an efficient choice of $J(\alpha;\cdot,\cdot)$ is the derivative function of $-U(\alpha;\cdot,\cdot)$ with respect to $\alpha$. We thus choose $J(\alpha;\cdot,\cdot)$ as the derivative function of $-U(\alpha;\cdot,\cdot)$ from now on.

Obviously, the above obeys the rule of RWS defined in Introduction. Moreover, the estimating function in (\ref{(estimating equation-1)}) is the weighted sum of the difference $ \alpha(x)-\widehat \alpha_{k-1}(x)$ and the unbiased estimating function $U(\alpha(x);\cdot,\cdot)$ in (\ref{(estimating equation)}). Therefore, the online updating estimator $\widehat \alpha_k(x)$ should be close to the previous estimator $\widehat \alpha_{k-1}(x)$ and should approximately satisfy the current estimating equation in (\ref{(estimating equation)}). Furthermore, in Remark 1 given below, we will show that actually the incremental updating estimating function in (\ref{(estimating equation-1)}) is an online updating score function if $U(\alpha(x);y,x)$ is a sore function from a likelihood.

Particularly, for the case of $U(\alpha(x);y,x)=(y-\alpha(x))$, solving equation (\ref{(estimating equation-1)}) results in the online updating estimator having the following closed representation:
\begin{eqnarray}\label{(solution)}\widehat \alpha_k(x)=\frac{\widehat \alpha_{k-1}(x)\sum_{j=1}^{k-1}\sum_{i\in {\bm i}_j}K_{h_j}(X_i-x)+\sum_{i\in {\bm i}_k}Y_iK_{h_k}(X_i-x)}{\sum_{j=1}^{k-1}\sum_{i\in {\bm i}_j} K_{h_j}(X_i-x)+\sum_{i\in {\bm i}_k} K_{h_k}(X_i-x)}.\end{eqnarray} Formally, $\widehat \alpha_k(x)$ can be expressed as
\begin{eqnarray}\label{(NW-solution)}\widehat \alpha_k(x)=\frac{\sum_{j=1}^k\sum_{i\in {\bm i}_j}Y_iK_{h_j}(X_i-x)}{\sum_{j=1}^k\sum_{i\in {\bm i}_j} K_{h_j}(X_i-x)}.\end{eqnarray} This implies that, similar to the N-W estimator in (\ref{(regression-estimator)}), the online updating estimator (\ref{(solution)}) is a N-W estimator as in (\ref{(NW-solution)}) with bandwidth $h_j$ depending the size of subset ${\bm d}_j$. For general case, solving equation (\ref{(estimating equation-1)})
may be easily done by the following incremental iterative algorithm:
\begin{eqnarray}\label{(iterative)}\widehat\alpha^{(s+1)}_k(x)=\widehat\alpha^{(s)}_k(x)+\left (\widehat{\bf J}_{k-1}+\sum_{i\in {\bm i}_k}J(\widehat \alpha^{(s)}_k(x);Y_i,X_i)K_{h_k}(X_i-x)\right)^{-1} U^{(s)}_k, \end{eqnarray} where
$\widehat{\bf J}_{k-1}=\sum_{j=1}^{k-1}\sum_{i\in {\bm i}_j}J(\widehat\alpha_j(x);Y_i,X_i) K_{h_j}(X_i-x)$, the initial one is chosen as $\widehat{\bf J}_0\equiv0$ by convention, and
\begin{eqnarray*}U^{(s)}_k&=&\widehat{\bf J}_{k-1}( \widehat\alpha^{(s)}_k(x) -\widehat \alpha_{k-1}(x))-
\sum_{i\in {\bm i}_k}U(\widehat\alpha^{(s)}_k(x),Y_i,X_i)K_{h_k}(X_i-x).\end{eqnarray*}

In the above estimation procedures, the method for bandwidth selection can be the classical ones, for example, the classical CV rule. The theoretical optimal choice of bandwidth will be given in the next
section, and the empirical choice will be discussed in simulation study.

Similarly, we can use the following incremental updating optimization method to construct the online updating estimator:
\begin{eqnarray}\label{(optimization)}\nonumber&&\min_{\alpha(x)}\sum_{j=1}^{k-1}\sum_{i\in {\bm i}_j}J(\widehat \alpha_j(x);Y_i,X_i) K_{h_j}(X_i-x)( \alpha(x)-\widehat \alpha_{k-1}(x))^2\\ && \ \ \ \ \ +
\sum_{i\in {\bm i}_k}L(\alpha(x);Y_i,X_i)K_{h_k}(X_i-x), \end{eqnarray} where $L(\alpha;y,x)$ is a given loss function and $J(\alpha;y,x)$ is a known weight function. An efficient choice of $J(\alpha;\cdot,\cdot)$ is the second-order derivative function of $L(\alpha;\cdot,\cdot)$ with respect to $\alpha$. In Remark 1 below, we will show the RWS in (\ref{(optimization)}) is in fact an online updating likelihood if $L(\alpha;y,x)$ is a likelihood. For the optimization problem, the iterative algorithm is similar to  (\ref{(iterative)}); the details are omitted here.

Generally, we can define the generic loss function and estimating equation as in (\ref{(intro-0)}) and (\ref{(intro-0-1)}) for the above nonparametric model. Because of the complexity in algorithm and theory, the details are omitted in this paper.

\subsection{Online updating likelihoods}

The weighted sums in (\ref{(estimating equation-1)}) and (\ref{(optimization)}) are constructed according to the motivating examples in (\ref{(updating-1)}), (\ref{(updating-2)}) and (\ref{(updating-3)}). We now explain the theoretical reasonability.
It can be verified that the methods (\ref{(estimating equation-1)}) and (\ref{(optimization)}) are equivalent under some regularity conditions. Thus, for conveniently explaining the theoretical reasonability of the RWS, we first consider the incremental updating optimization function in (\ref{(optimization)}) with $L(\alpha;y,x)$ being negative log likelihood and $k=2$. In this case, the key is to explain why we use the square loss $(\widehat\alpha_1(x)-\alpha(x))^2$ together with weight function $J(\alpha(x);y,x)$ in the first part of the optimization function. It is known that usually $\widehat\alpha_1(x)-\alpha^0(x)$ is normally distributed, asymptotically, with mean zero and the variance proportional to $1/E(J(\alpha^0(x);Y,X))$. Then, the negative (local) log likelihood derived from the asymptotic distribution of $\widehat\alpha_1(x)$ is equal to
$$\sum_{i\in {\bm i}_1}J(\widehat \alpha_1(x);Y_i,X_i) K_{h_1}(X_i-x)(\widehat\alpha_1(x)-\alpha(x))^2,$$ a weighted square loss function. By this resultant likelihood combined with the original likelihood $L(\alpha;y,x)$, we get the online updating (local) likelihood as
\begin{eqnarray}\label{(explanation)}&&\nonumber\sum_{i\in {\bm i}_1}J(\widehat \alpha_1(x);Y_i,X_i) K_{h_1}(X_i-x)( \alpha(x)-\widehat \alpha_1(x))^2\\&& +
\sum_{i\in {\bm i}_2}L(\alpha(x);Y_i,X_i)K_{h_2}(X_i-x). \end{eqnarray} This is just the special case of the incremental updating optimization function in (\ref{(optimization)}). The discussion reveals the following truth:

\

\noindent{\bf Remark 1.}
\begin{itemize}\item[] {\it Actually the RWS in (\ref{(optimization)}) is an incremental updating likelihood function, and the RWS in (\ref{(estimating equation-1)}) is an incremental updating score function, if $L(\alpha;y,x)$ and $U(\alpha;y,x)$ are likelihood and score functions respectively.
}\end{itemize}
For the generic optimization framework as in (\ref{(intro-0)}), when $L(\cdot;\cdot)$ is chosen as a likelihood and $D(\cdot,\cdot)$ is the likelihood derived from the asymptotic distribution of $\widehat\alpha_{k-1}(x)-\alpha^0(x)$, we sill have the above likelihood and score rules. Therefore, the above frameworks in deed lay the foundations of likelihood and score theories for the analysis of streaming data sets. As a consequence, the classical theoretical properties can be successfully achieved (for details see the next section).

\subsection{Extensions}
Furthermore, the RWS is a unified framework, it can be applied to various types of estimations, not only nonparametric estimation, but also the estimations in the parametric and semiparametric models. The extension to parametric models is direct, obviously.
Because of the particularity of semiparametric models, we here briefly show how to extend the incremental updating estimating equation (\ref{(estimating equation-1)}) into semiparametric models. Suppose that in addition to an unknown nonparametric function $\alpha(x)$,  an unknown parameter $\beta$ is included in a semiparametric model. According to the semiparametric estimation of Li and Liang (2008) and the strategy in (\ref{(estimating equation-1)}), the online updating procedure at the $k$-th step needs to solve the following two equations. The first is the following vectorial nonparametric estimating equation:
\begin{eqnarray}\label{(semi-1)}\nonumber&&\sum_{j=1}^{k-1}\sum_{i\in {\bm i}_j}J_{\alpha,\beta}((\widehat \alpha_j(x),\widehat\beta_j);Y_i,X_i) K_{h_j}(X_i-x)\left( (\alpha(x),\beta)^T-(\widehat \alpha_{k-1}(x),\widehat \beta_{k-1})^T\right)\\ && -
\sum_{i\in {\bm i}_k}U((\alpha(x),\beta);Y_i,X_i)K_{h_k}(X_i-x)=0 \end{eqnarray} for $(\alpha(x),\beta)^T$, where $U((\alpha(x),\beta);y,x)$ is a $2$-dimensional vector of unbiased estimating functions, and $J_{\alpha,\beta}((\alpha(x),\beta);\cdot,\cdot)$ is the derivative matrix of $-U((\alpha(x),\beta);\cdot,\cdot)$ with respect to $(\alpha,\beta)^T$. By the estimator $\widehat\alpha_k(x)$ of $\alpha(x)$ obtained from the first equation (\ref{(semi-1)}), we solve the following parametric estimating equation:
\begin{eqnarray}\label{(semi-2)} \sum_{j=1}^{k-1}\sum_{i\in {\bm i}_j}J_\beta((\widehat \alpha_j(X_i),\widehat\beta_j);Y_i,X_i) ( \beta - \widehat \beta_{k-1} )  -
\sum_{i\in {\bm i}_k}U_2((\widehat\alpha_k(X_i),\beta);Y_i,X_i) =0 \end{eqnarray} for $ \beta $, where $U_2((\alpha(x),\beta);y,x)$ is the 2-th element of the vector $U((\alpha(x),\beta);y,x)$, and $J_\beta((\alpha(x),\beta);\cdot,\cdot)$ is the derivative of $U((\alpha(x),\beta);y,x)$ with respect to $\beta$. Denote by $\widehat\beta_k$ the solution of $\beta$ obtained by the second equation (\ref{(semi-2)}). Then, the final solution at the $k$-th step updating is $(\widehat\alpha_k(x),\widehat\beta_k)^T$.

\setcounter{equation}{0}
\section{Theoretical properties}

We now establish the estimation consistency and asymptotic normality for the proposed online
updating estimation, and show its adaptability to perpetual streaming data sets and its asymptotic equivalence to the oracle estimator obtained by entire data sets.

Due to the equivalency between (\ref{(estimating equation-1)}) and (\ref{(optimization)}), we only investigate the theoretical properties of the online updating estimator $\widehat\alpha_k$ obtained by estimating equation (\ref{(estimating equation-1)}).
The theoretical properties obtained below can be extended into general cases such as spline estimation and local polynomial estimation and so on.

To the end, we need the following regularity conditions:
\begin{itemize}\item[(C1)] Kernel function $K(u)$ is symmetric with
respect to $u=0$, and satisfies $\int K(u)du=1$, $\int u^2
K(u)du<\infty$ and $\int u^2 K^2(u)du<\infty$.
\item[(C2)] Functions
$J(\alpha;y,x)$ and $U(\alpha;y,x)$
have the second-order continuous and bounded partial derivatives with respect to $\alpha$, $E(J(\alpha^0(X);Y,X)|X=x)\neq 0$ and $E(U^2(\alpha^0(X);Y,X)|X=x)$ exists for all $x\in [0,1]$, and the density function $f(x)$ of $X$ has second-order continuous derivative and satisfies $f(x)>0$ for all $x\in [0,1]$.
\item[(C3)] $|{\bm i}_j|\rightarrow\infty$, where $|{\bm i}_j|$ is the size of ${\bm i}_j$, i.e., the number of elements in ${\bm i}_j$, and all bandwidths satisfy $h_j\rightarrow 0$ and $h_j|{\cal I}_k|\rightarrow\infty$ for $j=1,\cdots,k$, where ${\cal I}_k=\bigcup_{j=1}^k{\bm i}_j$.
    \item[(C4)] The solution $\alpha^0(x)$ to the equation (\ref{(estimating equation)}) is unique, and has second-order continuous derivative for all $x\in [0,1]$.
\end{itemize}
Obviously, Conditions (C1)-(C3) are common for nonparametric kernel estimation (see, i.e., H\"{a}rdle, et al., 2004), and the assumption on the unique solution in Condition (C4) is also commonly used in the theory of estimating equation (see, i.e., Raymond et al., 1998).

\

\noindent{\bf Lemma 1.} {\it Under Conditions (C1)-(C4), the online updating estimator $\widehat\alpha_k(x)$ (i.e., the solution to the equation (\ref{(estimating equation-1)})) is consistent in probability for each $k$. }

The proof of the lemma is presented in Supplementary Materials. By the consistency,
we have the following Theorem.

\

\noindent{\bf Theorem 1.} {\it Under Conditions (C1)-(C4), the online updating estimator satisfies $\widehat\alpha_k(x)-\alpha^0(x)=O_p(\overline\delta_k)$  for $x\in (0,1)$ and each $k$, where $\overline\delta_k=\frac{1}{|{\cal I}_k|}\sum_{j=1}^k|{\bm i}_j|\delta_{jk}$ with $\delta_{jk}=O(h^2_j+1/\sqrt{h_j|{\cal I}_k|})$. }

In the theorem, the condition $x\in (0,1)$ is not a necessary constraint; that is, we use it only for a simple presentation, without boundary effect. The proof of the theorem is given in Supplementary Materials as well. For the theorem, we have the following explanations.

\

\noindent{\bf Remark 2. The standard convergence rate and the optimal bandwidth.} {\it
\begin{enumerate}
\item[1)] For checking the standard convergence rate, we consider the simple case when the sizes $|{\bm i}_j|$ are equal for all $j$, and then the bandwidths $h_j$ for all $j$ are equal to each other, denoted by $h=h_j$. In this case, the convergence rate of the online updating estimator $\widehat\alpha_k(x)$ is of order $O_p(\overline\delta_k)=O_p(h^2+1/\sqrt{h|{\cal I}_k|})$, the standard convergence rate of nonparametric kernel estimator computed on the entire data set ${\cal D}_k=\bigcup_{j=1}^k{\bm d}_j$.
    \item[2)] The theoretical optimal bandwidth, denoted by $h_j^*$, is of order $O(|{\cal I}_k|^{-1/5})$ for all $j$. It shows that the choice of the bandwidth $h_j$ should be much smaller than those in local estimators computed on local data sets ${\bm d}_j$, because $|{\bm d}_j|$ is much smaller than $|{\cal I}_k|$ and the number $k$ of the data sets grows infinitely with the observation time. It is difficult or impossible to achieve the optimal bandwidth unless the terminal time of streaming data sets is predetermined. We then call $h_j^*$ as oracle bandwidth as if the terminal time of streaming data sets was known in advance. This is the essential difference from the methods for parametric models with streaming data sets. The issue will be further discussed in simulation study.
    \end{enumerate}}

Denote $g(x)=\alpha''(x)+2\frac{\alpha'(x)f'(x)}{f(x)}$, $\|K\|_2=\int K^2(u)du$ and $\mu_2(K)=\int u^2K(u)du$.
Based on Lemma 1 and Theorem 1,
we can establish the asymptotic normality as stated in the following theorem.

\

\noindent{\bf Theorem 2.} {\it Under Conditions (C1)-(C4), each online updating estimator $\widehat\alpha_k(x)$ has the following asymptotic normality:
\begin{eqnarray*}\overline \upsilon_k^{-1}(\widehat\alpha_k(x)-\alpha^0(x)-\overline b_k(x))
\stackrel{d}\rightarrow N\left(0,\frac{E(U^2(\alpha^0;Y,X))\|K\|_2^2}{f(x)E^2(J(\alpha^0;Y,X))}\right) \ \mbox{ for } x\in (0,1), \end{eqnarray*} where $ \overline \upsilon_k=\frac{1}{|{\cal I}_k|}\sum_{j=1}^k|{\bm i}_j|\upsilon_{jk}$ with $\upsilon_{jk}=1/\sqrt{h_j|{\cal I}_k|}$, $\overline b_k(x)=\frac{g(x)\mu_2(K)}{2|{\cal I}_k|}\sum_{j=1}^k|{\bm i}_j|h^2_j$, and the notation $``\stackrel{d}\rightarrow"$ stands for convergence in distribution.
Particulary,
if $h_j=o(|{\cal I}_k|^{-1/5})$, then,
\begin{eqnarray*}\overline \upsilon_k^{-1}(\widehat\alpha_k(x)-\alpha^0(x))
\stackrel{d}\rightarrow N\left(0,\frac{E(U^2(\alpha^0;Y,X))\|K\|_2^2}{f(x)E^2(J(\alpha^0;Y,X))}\right) \ \mbox{ for } x\in (0,1). \end{eqnarray*}
}

The proof of the theorem is also given in Supplementary Materials. In the second conclusion of the theorem, we need the condition $h_j=o(|{\cal I}_k|^{-1/5})$ only for eliminating the asymptotic bias $\overline b_k(x)$.
From the theorem, we have the following findings.

\

\noindent{\bf Remark 3. Efficiency, adaptability and the oracle property.} {\it \begin{enumerate}
\item[1)] (Estimating efficiency) It can be easily proven that for the case where $U(\alpha;y,x)$ is the score function from a likelihood, $E(U^2(\alpha^0;Y,X))=E(J(\alpha^0;Y,X))$ under some regularity conditions. With the result, the asymptotic variance is equal to $\frac{\|K\|_2^2}{f(x)E(J(\alpha^0;Y,X))}$, implying the estimation efficiency.
\item[2)] (Adaptability to perpetual streaming
data sets) The convergence and asymptotic normality of the online updating estimator $\widehat\alpha_k(x)$ always hold for any $k$, without the constraint $k=O(n^c)$. This implies that the newly proposed method is adaptive to the situation where streaming data sets arrive perpetually with $k\rightarrow\infty$.
     \item[3)] (Oracle property) Let $\alpha^*(x)$ denote the oracle estimator, namely, it is the solution to the entire data estimating equation:
\begin{eqnarray*}\sum_{j=1}^k\sum_{i\in {\bm i}_j}
U(\alpha(x);Y_i,X_i)K_{h_k}(X_i-x)=0 \end{eqnarray*} for $\alpha(x)$. It can be easily proven by the theory of local estimating equation
(see, e.g., Carroll, et al., 1998) that
\begin{eqnarray*}\overline \upsilon_k^{-1}(\alpha^*(x)-\alpha^0(x)-\overline b_k(x))
\stackrel{d}\rightarrow N\left(0,\frac{E(U^2(\alpha^0;Y,X))\|K\|_2^2}{f(x)E^2(J(\alpha^0;Y,X))}\right) \ \mbox{ for } x\in(0,1). \end{eqnarray*}\end{enumerate} It indicates that the online updating estimator $\widehat\alpha_k(x)$ achieves the oracle property in the sense that it has the same behavior as the oracle estimator $\alpha^*(x)$, asymptotically.  }

\setcounter{equation}{0}
\section{Numerical analyses}

\subsection{Empirical evidences}

In this subsection, we provide the main results of simulation studies. For the further simulation results on  Cubic Spline estimation, see Supplemental Materials.

In the following, we evaluate our online updating approach through simulation studies under the following three typical models: the homoscedastic mean regression model, heteroscedastic mean regression model and conditional law model. We set that the full dataset consists of $n$ observations and $k$ batches ${\bm d}_j,j=1,\cdots,k$. According to the previous notation, the batch size of ${\bm d}_j$ is denoted by $|{\bm i}_j|$ for $j=1,\cdots,k$. To evaluate the effect of  sample size $n$ and batch size $|{\bm i}_j|$, as in Lou et al. (2020), we generate the datasets in the two ways by fixing one of the two parameters and varying another. The estimation performance is measured with the mean integrated squared error (MISE) derived by  200 replications. All the kernel estimators are constructed by the Gaussian kernel.

\subsubsection{ Homoscedastic mean regression model}

The homoscedastic mean regression model is formulated as
\begin{equation}
\label{model:1}
Y=\sin(2X)+\varepsilon,
\end{equation}
where $X\sim U[-3,3]$ is a one-dimensional covariate and $\varepsilon$ has a $N(0,0.2^2)$ law. In our RWS estimation procedure, we set the estimating function as $U(\alpha(x);y,x)=y-\alpha(x)$. For a comprehensive comparison, we consider the following estimators:

1) Our online updating estimator with full data bandwidth $h_f$, denoted by RWS$_{h_f}$.

2) Our online updating estimator with online updating bandwidth $h_k$, denoted by RWS$_{h_k}$.

3) The full data N-W estimator with full data bandwidth $h_f$, denoted by NWE$_f$.

4) The simple average of each batch N-W estimators, denoted by NWE$_a$.

In the above, the full data N-W estimator and simple average estimator are constructed by full dataset as if the full dataset were given in advance, and the full data bandwidth is chosen as $h_f=c_fn^{-1/5}$, where $c_f$ is determined by Cross-Validation criterion from the full dataset. For our RMS estimator, the online updating bandwidth is chosen as $h_k=c_1(\sum_{j=1}^k|{\bm i}|_j)^{-1/5}$ when the $k$-th data batch arrives, where $c_1$ is determined by Cross-Validation computed on the first batch.

The simulation results are reported in Table \ref{tab:1} and Table \ref{tab:2}. We have the following findings:

1) Our RWS$_{h_k}$ is much better than NWE$_a$ in the sense that the MISE of RWS$_{h_k}$ is significantly smaller than that of NWE$_a$, and moreover, our RWS$_{h_k}$ behaves in the same way as that of the NWE$_f$ when the sample size $n$ is large enough.

2) Our RWS$_{h_f}$ has the same performance as NWE$_f$. Actually, by (\ref{(NW-solution)}), the two estimators are equal if they share the same bandwidth.

3) Our RWS$_{f}$ and RWS$_{h_k}$ are robust to the  varying (from 30 to 1000)  of batch size $|{\bm i}|_j$, as the sample size is  fixed as $n=12000$, and the difference between RWS$_{f}$ and RWS$_{h_k}$ is not significant.

The above simulation results can illustrate the theoretical conclusions proposed in the previous section.

\begin{table}[htbp]
\centering
\caption{\small The MISE of different estimators under model (\ref{model:1}) are summarized over 200 replication, with fixed $n=12000$ and varying batch sizes $|{\bm i}|_j$.
}
\label{tab:1}
\begin{tabular}{c|c|c|c|c }
\hline
$|{\bm i}|_j$ & NWE$_f$ & NWE$_a$ & RWS$_{h_f}$ & RWS$_{h_k}$  \\
\hline
30 & $2.186\times 10^{-4}$ & $2.981\times 10^{-2}$   & $2.186\times 10^{-4}$ &$2.201\times 10^{-4}$ \\
\hline
50 & $2.186\times 10^{-4}$ & $1.384\times 10^{-2}$ & $2.186\times 10^{-4}$ & $2.287\times 10^{-4}$   \\
\hline
100 & $2.186\times 10^{-4}$ & $4.063\times 10^{-3}$ & $2.186\times 10^{-4}$ & $2.276\times 10^{-4}$   \\
\hline
300 & $2.186\times 10^{-4}$ & $8.117\times 10^{-4}$ & $2.186\times 10^{-4}$ & $2.273\times 10^{-4}$  \\
\hline
500 & $2.186\times 10^{-4}$ & $5.644\times 10^{-4}$ & $2.186\times 10^{-4}$ & $2.268\times 10^{-4}$  \\
\hline
1000 & $2.186\times 10^{-4}$ & $4.115\times 10^{-4}$ & $2.186\times 10^{-4}$ & $2.263\times 10^{-4}$  \\
\hline
\end{tabular}
\end{table}

\begin{table}[htbp]
\centering
\caption{\small The MISE of different estimators under model (\ref{model:1}) are summarized over 200 replication, with fixed batch size $|{\bm i}|_j=100$ and varying $n$.
}
\label{tab:2}
\begin{tabular}{ c|c|c|c|c }
\hline
$n$ & NWE$_f$ & NWE$_a$ & RWS$_{h_f}$ & RWS$_{h_k}$  \\
\hline
1000 & $1.308\times 10^{-3}$ & $5.734\times 10^{-3}$ & $1.308\times 10^{-3}$ & $1.395\times 10^{-3}$ \\
\hline
10000 & $2.439\times 10^{-4}$ & $4.123\times 10^{-3}$ & $2.439\times 10^{-4}$ & $2.456\times 10^{-4}$  \\
\hline
100000 & $6.432\times 10^{-5}$ & $3.938\times 10^{-3}$ & $6.432\times 10^{-5}$ & $6.437\times 10^{-5}$  \\
\hline
1000000 & $2.106\times 10^{-5}$ & $3.926\times 10^{-3}$ & $2.106\times 10^{-5}$ & $2.107\times 10^{-5}$  \\
\hline
\end{tabular}
\end{table}

\subsubsection{Heteroscedastic mean regression model}
Inspired by the empirical study in Chown \& M{\"u}ller (2018), we consider the following heteroscedastic mean regression setup:
\begin{equation}
\label{model:2}
Y=X+\cos(\pi X)+\left(\exp(X)-0.25\right)\varepsilon,
\end{equation}
where $X\sim U[-1,1]$ and $\varepsilon\sim N(0,1)$. In this example, we have a two-dimensional vector-valued estimating function $U\left(\alpha(x);y,x\right)=
\left(
\begin{array}{cccc}
y-r(x)\\
\left(y-r(x)\right)^2-\sigma^2(x)\\
\end{array}
\right )$,
where $\alpha(x)=\left(r(x),\sigma^2(x)\right)^T$, $r(x)=E(y|x)$ and $\sigma^2(x)=Var(y|x)$. The performances of four estimators RWS$_{h_f}$, RWS$_{h_k}$, NWE$_f$ and NWE$_a$ are reported in Table \ref{tab:3}  and Table \ref{tab:4}. Similar to the findings obtained in Example 1, for the above heteroscedastic model and the related estimators of $r(x)$ and $\sigma^2(x)$, we have the following findings:
\begin{itemize}
\item [1)] Our estimators by RWS$_{h_k}$ are much better than the estimators by NWE$_a$ because the MISEs of the estimators by RWS$_{h_k}$ are significantly smaller than those by NWE$_a$. Moreover, our estimators by RWS$_{h_k}$ have the similar behavior to those by NWE$_f$ when the sample size $n$ is large enough.
\item [2)] Our estimators by RWS$_{h_f}$ and RWS$_{h_k}$ are robust to the varying of batch size $|\bm{i}|_j$, as the sample size is fixed as $n=12000$.
\item [3)]
    In theory, our estimators by RWS$_{h_f}$ and NWE$_f$ should provide the same estimate for  $Var(Y|X)$, if they share the same bandwidth. Actually, the estimates are slightly different because  RWS$_{h_f}$ involves the online updating estimate of $E(Y|X)$ with partial data (instead of full data).

\end{itemize}

\begin{table}[htbp]
\centering
\caption{\small The MISE of different estimators under model (\ref{model:2}) are summarized over 200 replications, with fixed $n=12000$ and varying batch sizes $|{\bm i}|_j$.
}
\label{tab:3}
\begin{tabular}{c|c|c|c|c}
\hline
\multicolumn{5}{c}{$E(Y|X)$}\\
\hline
$|{\bm i}|_j$ & NWE$_f$ & NWE$_a$ & RWS$_{h_f}$ & RWS$_{h_k}$ \\
\hline
30 &   $3.353\times 10^{-3}$ & $2.495\times 10^{-2}$ & $3.353\times 10^{-3}$ & $3.504\times 10^{-3}$  \\
\hline
50 &   $3.353\times 10^{-3}$ & $1.626\times 10^{-2}$ & $3.353\times 10^{-3}$ & $3.500\times 10^{-3}$ \\
\hline
100 &  $3.353\times 10^{-3}$ & $9.997\times 10^{-2}$ & $3.353\times 10^{-3}$ & $3.495\times 10^{-3}$ \\
\hline
300 &  $3.353\times 10^{-3}$ & $5.485\times 10^{-3}$ & $3.353\times 10^{-3}$ & $3.481\times 10^{-3}$ \\
\hline
500 &  $3.353\times 10^{-3}$ & $4.478\times 10^{-3}$ & $3.353\times 10^{-3}$ & $3.474\times 10^{-3}$ \\
\hline
1000 & $3.353\times 10^{-3}$ & $3.713\times 10^{-3}$ & $3.353\times 10^{-3}$ & $3.471\times 10^{-3}$ \\
\hline
\multicolumn{5}{c}{$Var(Y|X)$}\\
\hline
30 &  $5.453\times 10^{-2}$ & $1.143\times 10^{0}$  & $6.006\times 10^{-2}$ & $6.606\times 10^{-2}$  \\
\hline
50 &  $5.453\times 10^{-2}$ & $6.693\times 10^{-1}$ & $5.969\times 10^{-2}$ & $6.565\times 10^{-2}$ \\
\hline
100 & $5.453\times 10^{-2}$ & $3.511\times 10^{-1}$ & $5.934\times 10^{-2}$ & $6.504\times 10^{-2}$ \\
\hline
300 & $5.453\times 10^{-2}$ & $1.633\times 10^{-1}$ & $5.841\times 10^{-2}$ & $6.352\times 10^{-2}$ \\
\hline
500 & $5.453\times 10^{-2}$ & $1.232\times 10^{-1}$ & $5.808\times 10^{-2}$ & $6.261\times 10^{-2}$ \\
\hline
1000 &$5.453\times 10^{-2}$ & $9.096\times 10^{-2}$ & $5.743\times 10^{-2}$ & $6.111\times 10^{-2}$ \\
\hline
\end{tabular}
\end{table}

\begin{table}[htbp]
\centering
\caption{\small The MISE of different estimators under model (\ref{model:2})  are summarized over 200 replications, with fixed batch size $|{\bm i}|_j=100$ and varying $n$.
}
\label{tab:4}
\begin{tabular}{c|c|c|c|c}
\hline
\multicolumn{5}{c}{$E(Y|X)$}\\
\hline
$n$ & NWE$_f$ & NWE$_a$ & RWS$_{h_f}$ & RWS$_{h_k}$ \\
\hline
1000 &    $2.859\times 10^{-2}$ & $2.690\times 10^{-2}$ & $2.859\times 10^{-2}$ & $2.664\times 10^{-2}$ \\
\hline
10000 &   $4.428\times 10^{-3}$ & $2.515\times 10^{-2}$ & $4.428\times 10^{-3}$ & $4.507\times 10^{-3}$ \\
\hline
100000 &  $8.190\times 10^{-4}$ & $2.422\times 10^{-2}$ & $8.190\times 10^{-4}$ & $9.510\times 10^{-4}$ \\
\hline
1000000 & $1.440\times 10^{-4}$ & $2.413\times 10^{-2}$ & $1.440\times 10^{-4}$ & $2.140\times 10^{-4}$ \\
\hline
\multicolumn{5}{c}{$Var(Y|X)$}\\
\hline
1000    & $2.566\times 10^{-1}$ & $6.281\times 10^{-1}$ & $3.041\times 10^{-1}$ & $3.454\times 10^{-1}$ \\
\hline
10000   & $5.560\times 10^{-2}$ & $3.449\times 10^{-1}$ & $6.409\times 10^{-2}$ & $7.977\times 10^{-2}$ \\
\hline
100000  & $1.533\times 10^{-2}$ & $3.377\times 10^{-1}$ & $1.661\times 10^{-2}$ & $2.354\times 10^{-2}$ \\
\hline
1000000 & $4.736\times 10^{-3}$ & $3.370\times 10^{-1}$ & $4.917\times 10^{-3}$ & $7.640\times 10^{-3}$ \\
\hline
\end{tabular}
\end{table}

\subsubsection{Conditional law model}
In this example, we consider the following conditional law model
\begin{equation}
\label{model:3}
Y\sim Gamma\left(\exp\left(\frac{\cos(X)}{2}\right),1\right),
\end{equation}
where the one-dimensional covariate $X\sim U[-1,1]$. Here we compare our methods RWS$_{h_f}$ and RWS$_{h_k}$ with two other methods: the full data-based nonparametric maximum likelihood estimator (NML$_f$) with full data bandwidth $h_f$ and the simple average of each batch nonparametric maximum likelihood estimator (NML$_a$). The bandwidths are selected by the same way as in model (\ref{model:1}).  To fit  this model,  as analyzed before, we adopt the score function of maximum likelihood as the estimating function $U\left(\alpha(x); y, x\right)$ and choose $J(a; y, x)$ as  the derivative function of  $-U(\alpha; y, x)$. Consider that $J(a; y, x)$ is a function of $x, y$, we use the incremental iterative algorithm (3.5) to construct our RWS estimators.   The simulation results  are shown in Table \ref{tab:5} and Table \ref{tab:6}. We have the following findings:
\begin{itemize}
\item [1)] The MISE of RWS$_{h_k}$ is significantly smaller than that of NML$_a$, implying that our RWS$_{h_k}$ is much better than NML$_a$. Moreover, our RWS$_{h_k}$ behaves in the same way as that of the NML$_f$ when the sample size $n$ is large enough.
\item [2)] For the fixed size $n$ of full sample, our RWS$_{h_f}$ and RWS$_{h_k}$ are robust to the choice of batch size $|\bm{i}|_j$.
\end{itemize}

\begin{table}[htbp]
\centering
\caption{\small The MISE of different estimators under model (\ref{model:3}) are summarized over 200 replications, with fixed $n=12000$ and varying batch sizes $|{\bm i}|_j$.
}
\label{tab:5}
\begin{tabular}{c|c|c|c|c}
\hline
$|{\bm i}|_j$ & NML$_f$ & NML$_a$ & RWS$_{h_f}$ & RWS$_{h_k}$ \\
\hline
30   & $1.314\times 10^{-3}$ & $1.662\times 10^{-2}$ & $1.341\times 10^{-3}$ & $2.390\times 10^{-3}$  \\
\hline
50   & $1.314\times 10^{-3}$ & $1.226\times 10^{-2}$ & $1.340\times 10^{-3}$ & $1.698\times 10^{-3}$ \\
\hline
100  & $1.314\times 10^{-3}$ & $8.408\times 10^{-3}$ & $1.338\times 10^{-3}$ & $1.692\times 10^{-3}$ \\
\hline
300  & $1.314\times 10^{-3}$ & $4.860\times 10^{-3}$ & $1.333\times 10^{-3}$ & $1.672\times 10^{-3}$ \\
\hline
500  & $1.314\times 10^{-3}$ & $3.826\times 10^{-3}$ & $1.325\times 10^{-3}$ & $1.656\times 10^{-3}$ \\
\hline
1000 & $1.314\times 10^{-3}$ & $2.824\times 10^{-3}$ & $1.317\times 10^{-3}$ & $1.624\times 10^{-3}$ \\
\hline
\end{tabular}
\end{table}

\begin{table}[htbp]
\centering
\caption{\small The MISE of different estimators under model (\ref{model:3}) are summarized over 200 replications, with fixed batch size $|{\bm i}|_j=100$ and varying $n$.
}
\label{tab:6}
\begin{tabular}{c|c|c|c|c}
\hline
$n$ & NML$_f$ & NML$_a$ & RWS$_{h_f}$ & RWS$_{h_k}$ \\
\hline
1000   & $9.770\times 10^{-3}$ & $1.294\times 10^{-2}$ & $9.790\times 10^{-3}$ & $9.835\times 10^{-3}$ \\
\hline
10000  & $1.871\times 10^{-3}$ & $8.833\times 10^{-3}$ & $1.883\times 10^{-3}$ & $2.098\times 10^{-3}$ \\
\hline
100000 & $4.300\times 10^{-4}$ & $8.281\times 10^{-3}$ & $4.370\times 10^{-4}$ & $5.151\times 10^{-4}$ \\
\hline
1000000& $1.140\times 10^{-4}$ & $8.235\times 10^{-3}$ & $1.166\times 10^{-4}$ & $1.420\times 10^{-4}$ \\
\hline
\end{tabular}
\end{table}

\subsection{Real data application}

We  analyze the Air Quality dataset provided by  Vito et al. (2008), which is available in UCI Machine Learning  Repository\footnote{http://archive.ics.uci.edu/ml/datasets/Air+Quality}. The dataset contains the CO concentration provided by an air pollution monitoring station  and the readouts of a metal oxide chemical sensor which can be used to measure the CO concentration. The monitoring station could provide a true  CO concentration value,  while the sensor could output a value correlated but unequal to the true CO concentration value. Despite such defect, the sensors are preferred in practice because they are  low cost and easy to deploy, helpful for raising the density of monitoring networks. So, our goal is to model the correlation between the sensor data and CO concentration.   The CO concentration was measured hourly  from March 2004 to April 2005, resulting in 9358 observations. We delete the days that suffer from serious data missing, and then obtain a data steam of 303 days,  each day including  16 $\sim$ 24 observations. Here we take the data of each day as a batch, implying that the data stream consists of 303 batches.

\begin{figure}
\centering
\includegraphics[width=10cm]{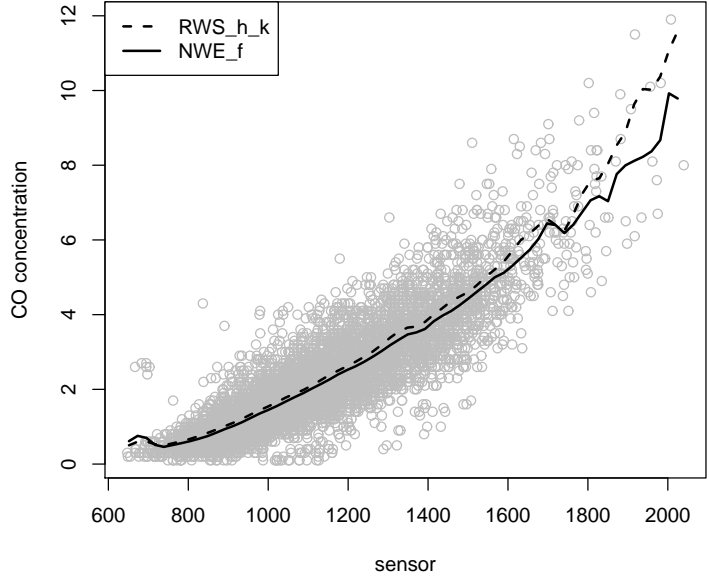}
\caption{\small The regression curves}
\label{fig:1}
\end{figure}

\begin{figure}
\centering
\includegraphics[width=10cm]{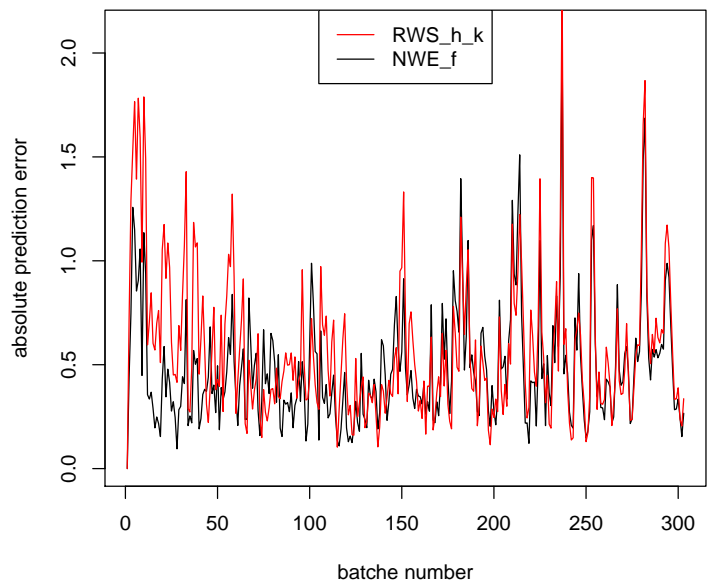}
\caption{\small The absolute prediction error}
\label{fig:2}
\end{figure}

\begin{figure}
\centering
\includegraphics[width=10cm]{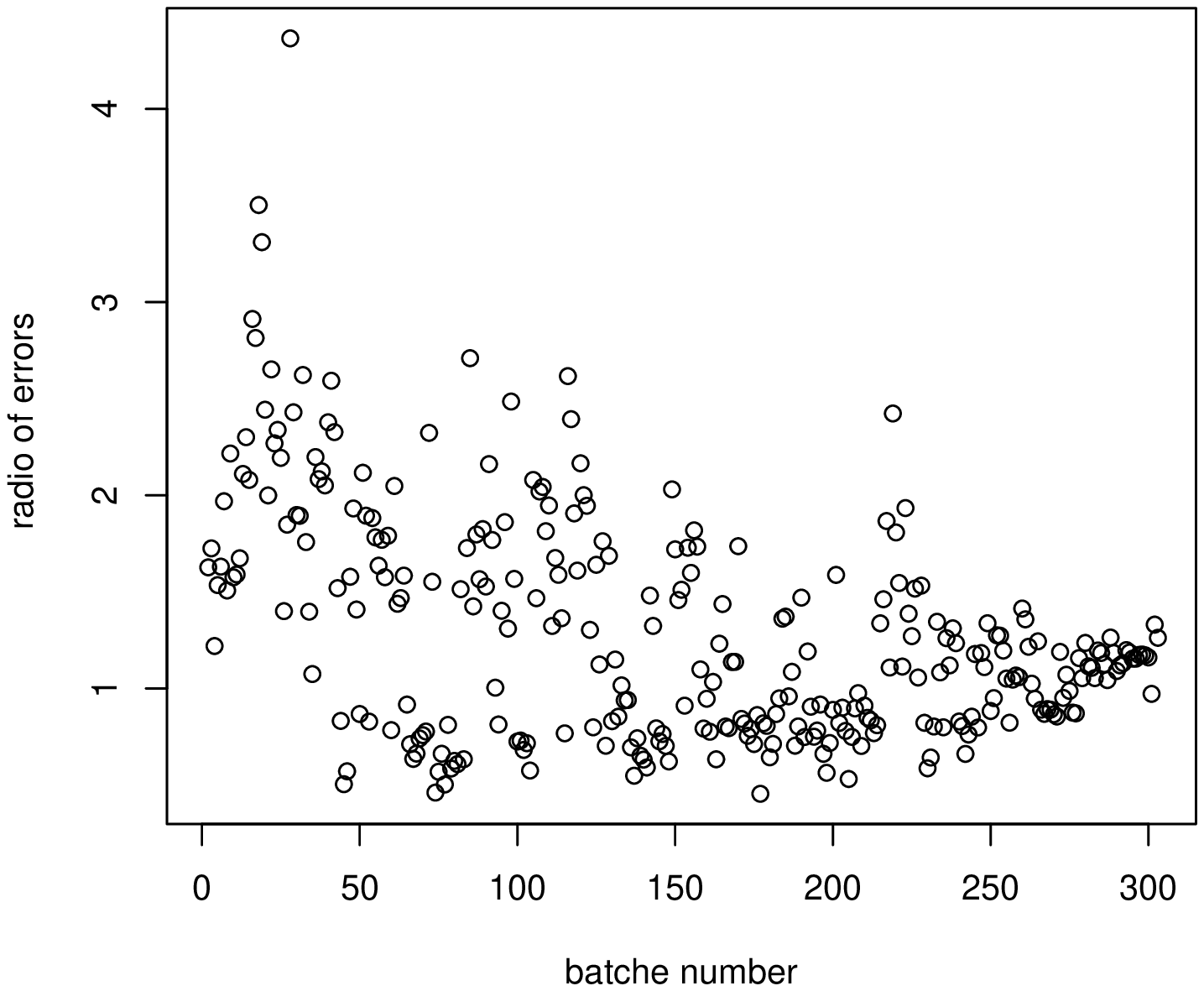}
\caption{\small The ration of the average absolute prediction errors between RWS$_{h_f}$ and NWE$_f$.}
\label{fig:3}
\end{figure}

Fig. \ref{fig:1} depicts the estimated regression curves of our RWS$_{h_k}$ with online updating bandwidth $h_k$, and  full data N-W estimator NWE$_f$ with full data bandwidth $h_f$. It is seen that the curves of two methods are close to each other in most cases,  except for  several imperfect cases in  the top right corner where the data are sparse.
In Fig. \ref{fig:2},  we show the average absolute prediction errors of RWS$_{h_k}$ and NWE$_f$ on the $k$-th batch. Recall that the RWS$_{h_k}$ only employs the $(k-1)$th data batch to predict the value on th $k$-th data batch, while the NWE$_f$ uses all the $k-1$ data batches to predict the value on th $k$-th data batch. Even so, Fig. \ref{fig:2} illustrates that the varying trends of errors the two methods are basically consistent. Further, we evaluate the radio of the average absolute prediction errors of two methods in Fig \ref{fig:3}. It is seen that the value tends to one with the increasing of batch number, implying that the performance of RMS$_{h_k}$ is similar to that of NWE$_f$ as the sample size is sufficiently large.

\begin{figure}
\centering
\includegraphics[width=10cm]{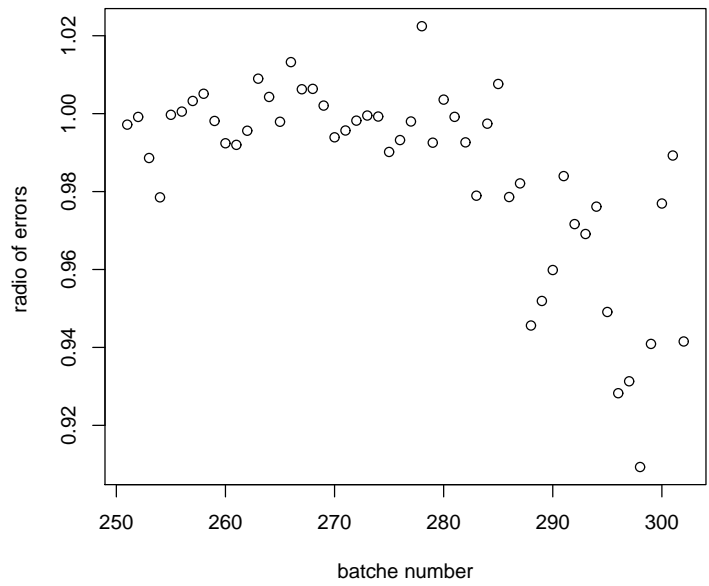}
\caption{\small The ration of the average absolute prediction errors between RWS$_{h_f}$ and NWE$_p$.}
\label{fig:4}
\end{figure}

Furthermore, we compare our RWS$_{h_k}$ with the  N-W estimator  with partial data for training (denoted as NWE$_p$). For such an estimator,  we use the first 250 data  batches for training and the remaining 53 batches for predicting.  Fig. \ref{fig:4}  depicts the ratio of the average absolute prediction errors of RWS$_{h_k}$ and NWE$_p$, which tends to decrease with the new batches arriving. This implies that our online updating method could provide more timely and  accurate predictions than the traditional NWE$_p$ method.

\newpage

\setcounter{equation}{0}
\section{Conclusions and future works}

It was shown in Introduction that although a large number of statistical methods and computational recipes have been developed to address the challenge of analyzing the models with streaming data sets, the unified notion and strategy of online updating likelihood (or online updating loss function) and online updating estimating equation have not been built in the existing literature. To address these issues, unified frameworks of renewable weighted sums (RWS) were established in the previous sections for constructing various online updating estimations in the models with streaming data sets. It was verified in the previous sections that the newly defined RWS plays the role of online updating likelihood (or online updating loss function) and online updating estimating equation, and then founds the theoretical foundation for general online updating statistical inferences. Furthermore, the structure of the RWS is intuitive and heuristic, the algorithm is computationally simple, and the method applies to various type of models, such as parametric models, nonparametric models and semiparametric models.

Also it was stated in Introduction that another challenging issue in the area of streaming data sets is how to relax or remove the constraint on the number of streaming data sets. The previous section showed that the newly defined RWS can be free of the constraint, consequently, it is adaptive to the situation where streaming data sets arrive perpetually. Moreover, the online updating estimator by the RWS possesses estimation consistency, asymptotic normality and the oracle property. For the proposed kernel estimator in nonparametric models, the
optimal bandwidth was attained and the method for sequentially choosing bandwidth was suggested in the previous sections.  The behavior of the method was further illustrated by various numerical
examples from simulation experiments and real data analysis. The simulation verified that the finite performance of the new method is much better than the competitors and has the similar behavior as that of the oracle estimator.

For the kernel estimator in nonparametric models, however, the optimal bandwidth is an oracle choice as if the terminal time of streaming data sets was known in advance. Then, it is difficult or impossible to achieve the optimal bandwidth unless the terminal time of streaming data sets is predetermined. This is the essential difference from the methods for parametric models with streaming data sets. For the generic loss function as in (\ref{(intro-0)}), the algorithm and theory are complicated, the difficulty stems from the facts that the loss function $L_0(u,\cdot;\cdot)$ may not be differentiable and the corresponding estimating function may have no closed expression.
Furthermore, it is difficult to extend the proposed method into the strategy of divide-and-conquer in general models. These are interesting issues and are worth further study in the future.

\

\section*{References}
\begin{description}

\item Amari, S.-I., Park, H. and Fukumizu, K. (2000). Adaptive method of realizing natural gradient learning for multilayer perceptrons. {\it Neurl Computn}, {\bf 12}, 1399-1409.

\item Bordes, A., Bottou, L. and Gallinari, P. (2009). Sgd-qn: careful quasi-Newton stochastic gradient descent. {\it J. Mach. Learn. Res.}, {\bf 10}, 1737-1754.

\item Bucak, S. S. and Gunsel, B. (2009). Incremental subspace learning via non-negative matrix factorization. {\it Pattern Recognition}, {\bf 42}, 788-797.

\item Chen, X. and Xie, M.-G. (2014). A split-and-conquer approach for analysis of extraordinarily large data. {\it Statistica Sinica}, {\bf 24}, 1655-1684.

\item Chow, J. and M\"{u}ller,  U. (2018). Detecting heteroscedasticity in nonparametric regression using weighted empirical processes.  {\it J. R. Statist. Soc. B}, {\bf 80}, 951-974.

\item Duchi, J., Hazan, E. and Singer, Y. (2011). Adaptive subgradient methods for online learning and stochastic optimization. {\it J. Mach. Learn. Res}., {\bf 12}, 2121-2159.

\item H\"{a}rdle, W., M\"{u}ller, M., Sperlich, S. and Werwatz, A. (2004). {\it Nonparametric and semiparametric models.} Springer.

\item Hazan, E., Agarwal, A. andKale, S. (2007). Logarithmic regret algorithms for online convex optimization. {\it J. Mach. Learn. Res}., {\bf 69}, 169-192.

\item Hao, S., Zhao, P., Lu, J., Hoi, S. C. H., Miao, C. and Zhang, C. (2016). Soal: second-order online active learning. {\it Int. Conf. Data Mining, Barcelona}.

\item Kleiner, A., Talwalkar, A., Sarkar, P. and Jordan, M. I. (2014) A scalable bootstrap for massive data. {\it J. R. Statist}.
Soc. B, {\bf 76}, 795-816.

\item Li, R. and Liang, H. (2008). Variable selection in semiparametric regression modeling.
{\it Ann. Statist}. 36 261-286.

\item Liang, F., Cheng, Y., Song, Q., Park, J. and Yang, P. (2013). A resampling-based stochastic approximation method
for analysis of large geostatistical data. {\it J. Am. Statist. Ass}., {\bf 108}, 325-339.

\item Lin, N. and  Xi, R. (2011). Aggregated estimating equation estimation. {\it Statistics and Its Interface}, {\bf 4}, 73-83.

\item Lin, L. and Zhang, R. C. (2002). Three methods of empirical Euclidean likelihood for two samples and their comparison. {\it Chinese Journal of Applied Probability and Statistics}, {\bf 18}, 4, 393-399.

\item Liu, D. C. and Nocedal, J. (1989). On the limited memory bfgs method for large scale optimization.
{\it Mathematical Programming}, {\bf 45}, 503-528.

\item Luo, L. and Song, P. X. -K. (2020). Renewable estimation and incremental inference in
generalized linear models with streaming data sets. {\it J. R. Statist. Soc. B}, {\bf 82}, 69-97.

\item Ma, P., Mahoney, M.W. and Yu, B. (2015). A statistical perspective on algorithm leveraging. {\it J. Mach. Learn. Res}., {\bf 6}, 861-911.

\item Maclaurin, D.and Adams, R. P. (2014). Firefly Monte Carlo: Exact MCMC with subsets of data. arXiv preprint. arXiv:1403.5693.

\item Nadaraya, E. A. (1964). On estimating regression. {\it Theory of probability and its application}, {\bf 10}, 186-190.

\item Neiswanger,W.,Wang, C., and Xing, E. (2013). Asymptotically exact, embarrassingly parallel MCMC. arXiv
preprint. arXiv:1311.4780.

\item Nion, D. and Sidiropoulos, N. D. (2009). Adaptive algorithms to track the PARAFAC decomposition of a thirdorder tensor. {\it IEEE Trans. Signl Process}, {\bf 57}, 2299-2310.

\item Raymond, J. Carroll, R. J., Ruppert, D. and Welsh, A. (1998). Local estimating equations. {\it Journal of the American Statistical Association}, {\bf 93}, 214-227.

\item Nocedal, J. and Wright, S. J. (1999). {\it Numerical optimization}. Springer-Verlag, New York.

\item Pillonetto, G., Schenato, L. and Varagnolo, D. (2019). Distributed multi-agent gaussian regression
via finite-dimensional approximations. {\it IEEE Transactions on Pattern Analysis and Machine Intelligence,} {\bf 41}, No. 9, 2098-2111.

\item Robbins, H. and Monro, S. (1951). A stochastic approximation method. {\it Ann. Math. Statist}., {\bf 22}, 400-407.

\item Schifano, E. D., Wu, J., Wang, C., Yan, J. and Chen, M. H. (2016). Online updating of statistical inference in the
big data setting. {\it Technometrics}, {\bf 58}, 393-403.

\item Scott, S. L., Blocker, A. W., Bonassi, F. V., Chipman, H., George, E., and  McCulloch, R. (2013). {\it Bayes and Big Data: The Consensus Monte Carlo Algorithm, EFaBBayes 250 Conference}, 16.

\item Schraudolph, N. N., Yu, J. and G\"{u}nter, S. (2007).  A stochastic quasi-Newton method for online convex optimization. {\it Proc. Mach. Learn. Res}., 2, 436-443.

\item Song, Q. and Liang, F. (2014). A split-and-merge Bayesian variable selection approach for ultrahigh dimensional regression. {\it Journal of the Royal Statistical Society, Series B}, {\bf 77}, 947-972.

\item Toulis, P. and Airoldi, E. M. (2015). Scalable estimation strategies based on stochastic approximations: classical results and new insights. {\it Statist. Comput}., {\bf 25}, 781-795.

\item Vaits, N., Moroshko, E. and Crammer, K. (2015). Second-order non-stationary online learning for regression. {\it J. Mach. Learn. Res}., {\bf 16}, 1481-1517.

\item Wang, C., Chen, M. H., Wu, J., Yan, J., Zhang, Y. and
Schifan, E. (2018). Online updating method with new variables
for big data streams. {\it The Canadian Journal of Statistics}, {\bf 46}, 2018, 123-146.

\item Watson, G. S. (1964). Smooth regression analysis. {\it Sankhy\={a}, Series A}, {\bf 26}, 359-370.

\item Xue, Y.,  Wang, H., Yan, J. and Schifano, E. D. (2019). An online updating approach for testing the proportional
hazards assumption with streams of survival data. {\it Biometric}, (to appear).

\end{description}

\end{document}


\title{Supplement to ``Unified Rules of Renewable Weighted Sums for Various Online Updating Estimations"
}
\author{Lu Lin$^1$, Weiyu Li$^1$\footnote{The corresponding
author. Email: liweiyu@sdu.edu.cn. The research was
supported by NNSF projects (11971265) of China.} \ and Jun Lu$^2$
\\
\small $^1$Zhongtai Securities Institute for Financial Studies, Shandong University, Jinan, China\\
\small $^2$School of Statistics and Mathematics, Zhejiang Gongshang University, Hangzhou, China
 }
\date{}
\maketitle

\vspace{-0.7cm}

\baselineskip=20pt

\section*{S.1. Proofs}

{\it Proof of Lemma 1.} By Conditions (C1)-(C4) and the theory of estimating equation, it can be verified that $\widehat\alpha_1(x)$ is consistent in probability. We then suppose that $\widehat\alpha_1(x),\cdots, \widehat\alpha_{k-1}(x)$ are consistent. The remaining task is to prove the consistency of $\widehat\alpha_k(x)$. Denote
\begin{eqnarray*}&&\widehat{\bf J}_k=\sum_{j=1}^k\sum_{i\in {\bm i}_j}J(\widehat\alpha_j(x);Y_i,X_i) K_{h_j}(X_i-x),\\ &&{\bf J}_k(\alpha)=\sum_{j=1}^k\sum_{i\in {\bm i}_j}J( \alpha(x);Y_i,X_i) K_{h_j}(X_i-x),\\&& J_j(\alpha)=\sum_{i\in {\bm i}_j}J(\alpha(x);Y_i,X_i) K_{h_j}(X_i-x),\\ && U_j(\alpha)=\sum_{i\in {\bm i}_j}U(\alpha(x),Y_i,X_i)K_{h_j}(X_i-x).\end{eqnarray*}
Note that $\widehat\alpha_k(x)$ is the solution to equation (3.2). Then
\begin{eqnarray}\label{(proof-1)}\frac{1}{|{\cal I}_k|}\widehat{\bf J}_{k-1}( \widehat\alpha_k(x)-\widehat \alpha_{k-1}(x))-\frac{1}{|{\cal I}_k|}
U_k(\widehat\alpha_k)=0. \end{eqnarray} Under the conditions of the theorem and the consistency of $\widehat\alpha_1(x),\cdots, \widehat\alpha_{k-1}(x)$ , we have \begin{eqnarray}\label{(proof-2)}\nonumber&&\hspace{-3ex}\frac{1}{|{\cal I}_k|}\widehat{\bf J}_{k-1}=\frac{1}{|{\cal I}_k|}{\bf J}_{k-1}(\alpha^0)+o_p(1)\\\nonumber&&\hspace{-3ex}\frac{1}{|{\cal I}_k|}{\bf J}_{k-1}(\alpha^0)( \widehat\alpha_k(x)-\widehat \alpha_{k-1}(x))=\frac{1}{|{\cal I}_k|}{\bf J}_{k-1}(\alpha^0)( \widehat\alpha_k(x)- \alpha^0(x))+o_p(1),\\&& \hspace{-3ex}\nonumber\frac{1}{|{\cal I}_k|}
U_k(\widehat\alpha_k)= \frac{1}{|{\cal I}_k|}U_k(\alpha^0)+\frac{1}{|{\cal I}_k|}J_k(\alpha^0)(\widehat\alpha_k(x)-\alpha^0(x)) +O_p\left(\frac{|{\bm i}_k|}{|{\cal I}_k|}\|\widehat\alpha_k(x)-\alpha^0(x)\|^2\right),\\ &&\hspace{-3ex}
\frac{1}{|{\cal I}_k|}{\bf J}_k(\alpha^0)=E(J(\alpha^0(X);Y,X)|X=x)+o_p(1).\end{eqnarray}
Then, it follows from (\ref{(proof-1)}) and (\ref{(proof-2)}) that
\begin{eqnarray*}  E(J(\alpha^0(X);Y,X)|X=x)( \widehat\alpha_k(x)-  \alpha^0(x))+O_p\left(\frac{|{\bm i}_k|}{|{\cal I}_k|}\|\widehat\alpha_k(x)-\alpha^0(x)\|^2\right)=o_p(1), \end{eqnarray*} which implies the consistency of $\widehat\alpha_k(x)$. $\square$

\

\noindent{\it Proof of Theorem 1.} It can be proven by theory of nonparametric estimating equation (see, e.g., Carroll, et al., 1998) that $\widehat\alpha_1(x)-\alpha^0(x)=O_p(\delta_1)=O_p(\overline\delta_1)$. We then suppose that $\widehat\alpha_j(x)-\alpha^0(x)=O_p(\overline\delta_{j})$ for $j=1,\cdots,k-1$. The remaining task is to prove $\widehat\alpha_k(x)-\alpha^0(x)=O_p(\overline\delta_k)$.
By Lemma 1 and Taylor expansion, we have
\begin{eqnarray}\label{(Theorem2-1)}&&\nonumber\hspace{-5ex}\sum_{j=1}^kU_j(\alpha^0)\\&&\hspace{-5ex}
=\sum_{j=1}^k\left(U_j(\widehat\alpha_j(x))
+J_j(\widehat\alpha_j(x))(\widehat\alpha_j(x)-\alpha^0(x))+O(|{\bm i}_j|\|\widehat\alpha_j(x)-\alpha^0(x)\|^2)\right).\end{eqnarray} This equation and
(\ref{(proof-1)}) together lead to
\begin{eqnarray}\label{(Theorem2-2)}\sum_{j=1}^kU_j(\alpha^0)=U_1(\widehat\alpha_1(x))+\widehat{\bf J}_k(\widehat\alpha_k(x)-\alpha^0(x))+\sum_{j=1}^kO_p(|{\bm i}_j|\|\widehat\alpha_j(x)-\alpha^0(x)\|^2).\end{eqnarray}
Note that $$\frac{1}{|{\cal I}_k|}\sum_{j=1}^kU_j(\alpha^0)=O_p(\overline\delta_k)$$ by the theory of kernel estimation (see, i.e., H\"{a}rdle, et al., 2004) .
Then
\begin{eqnarray}\label{(Theorem2-3)}&&\nonumber\frac{1}{|{\cal I}_k|}\widehat{\bf J}_k(\widehat\alpha_k(x)-\alpha^0(x))+O_p\left(\frac{|{\bm i}_k|}{|{\cal I}_k|}\|\widehat\alpha_k(x)-\alpha^0(x)\|^2\right)\\&&=-\frac{1}{|{\cal I}_k|}U_1(\widehat\alpha_1(x))+\sum_{j=1}^{k-1}O_p\left(\frac{|{\bm i}_j|}{|{\cal I}_k|}\|\widehat\alpha_j(x)-\alpha^0(x)\|^2\right)+O_p\left(\overline\delta_k\right).\end{eqnarray}
It is supposed that $\widehat\alpha_j(x)-\alpha^0(x)=O_p(\overline\delta_{j})$ for $j=1,\cdots,k-1$. Thus, the asymptotic order of the term on the right-hand side of (\ref{(Theorem2-3)}) is of order $O_p(\overline\delta_k)$.
We then have $\widehat\alpha_k(x)-\alpha^0(x)=O_p(\overline\delta_k)$. $\square$

\

\noindent{\it Proof of Theorem 2.} By (\ref{(Theorem2-2)}) and the consistency of $\widehat\alpha_k(x)$, we have
\begin{eqnarray*}\nonumber&&\frac{1}{|{\cal I}_k|}{\bf J}_k(\alpha^0)(\widehat\alpha_k(x)-\alpha^0(x))+o_p(\widehat\alpha_k(x)-\alpha^0(x))\\&&=\frac{1}{|{\cal I}_k|}\sum_{j=1}^kU_j(\alpha^0)+\frac{1}{|{\cal I}_k|}\sum_{j=1}^kO_p(|{\bm i}_j|\|\widehat\alpha_j(x)-\alpha^0(x)\|^2). \end{eqnarray*}
It indicates that
\begin{eqnarray*}\frac{1}{|{\cal I}_k|}{\bf J}_k(\alpha^0)(\widehat\alpha_k(x)-\alpha^0(x))\stackrel{d}=\frac{1}{|{\cal I}_k|}\sum_{j=1}^kU_j(\alpha^0)+\frac{1}{|{\cal I}_k|}\sum_{j=1}^kO_p(|{\bm i}_j|\|\widehat\alpha_j(x)-\alpha^0(x)\|^2), \end{eqnarray*} where the notation $``\stackrel{d}="$ stands for ``equal in distribution". This equation and Theorem 1 result in
\begin{eqnarray}\label{(Theorem2-4)}\widehat\alpha_k(x)-\alpha^0(x)
\stackrel{d}=\left(\frac{1}{|{\cal I}_k|}{\bf J}_k(\alpha^0)\right)^{-1}\frac{1}{|{\cal I}_k|}\sum_{j=1}^kU_j(\alpha^0)+o_p(\overline \upsilon_k). \end{eqnarray}
It can be verified by nonparametric estimation theory (see, i.e., H\"{a}rdle, et al., 2004) that \begin{eqnarray*}&&E\left(\frac{1}{|{\bm i}_j|}U_j(\alpha^0)\right)=\frac{g(x)\mu_2(K)}{2}h^2_j+o_p(h^2_j), \\&&Var\left(\frac{1}{|{\bm i}_j|}U_j(\alpha^0)\right)=\frac{\upsilon^2_{jk}E(U^2(\alpha^0;Y,X))\|K\|_2^2}{f(x)}+o_p(\upsilon^2_{jk}).\end{eqnarray*}
By these results and the central limit theorem of kernel estimator (see, i.e., H\"{a}rdle, et al., 2004), we have
\begin{eqnarray}\label{(Theorem2-5)}\upsilon^{-1}_{jk}\left(\frac{1}{|{\bm i}_j|}U_j(\alpha^0)-\frac{g(x)\mu_2(K)}{2}h^2_j\right)\stackrel{d}\rightarrow N\left(0,\frac{E(U^2(\alpha^0;Y,X))\|K\|_2^2}{f(x)}\right).\end{eqnarray}
Moreover, it can verified by nonparametric estimation theory (see, i.e., H\"{a}rdle, et al., 2004) that \begin{eqnarray}\label{(Theorem2-6)}\frac{1}{|{\cal I}_k|}{\bf J}_k(\alpha^0)=E(J(\alpha^0;Y,X))+o_p(1).\end{eqnarray}
Then, by (\ref{(Theorem2-4)}), (\ref{(Theorem2-5)}), (\ref{(Theorem2-6)}) and the independence among ${\bm d}_j,j=1,\cdots,k$, we have
\begin{eqnarray*}\overline \upsilon_k^{-1}(\widehat\alpha_k(x)-\alpha^0(x)-\overline b_k(x))
\stackrel{d}\rightarrow N\left(0,\frac{E(U^2(\alpha^0;Y,X))\|K\|_2^2}{f(x)E^2(J(\alpha^0;Y,X))}\right). \end{eqnarray*} Particulary,
if $h_j=o(|{\cal I}_k|^{-1/5})$, then, $\upsilon_k^{-1}\overline b_k(x)=o_p(1)$. As a result,
\begin{eqnarray*}\overline \upsilon_k^{-1}(\widehat\alpha_k(x)-\alpha^0(x))
\stackrel{d}\rightarrow N\left(0,\frac{E(U^2(\alpha^0;Y,X))\|K\|_2^2}{f(x)E^2(J(\alpha^0;Y,X))}\right). \end{eqnarray*} The proof is completed.
$\square$

\

\noindent{\bf References}
\begin{description}

\item Raymond J. Carroll, R. J., Ruppert, D. and Welsh, A. (1998). Local estimating equations. {\it Journal of the American Statistical Association}, {\bf 93}, 214-227.

\item H\"{a}rdle, W., M\"{u}ller, M., Sperlich, S. and Werwatz, A. (2004). {\it Nonparametric and semiparametric models.} Springer.

\end{description}

\subsection*{S.2. Further simulation studies}

For generality, we further construct our RWS estimators via Cubic Spline estimation, and test their performance on the mean regression model (5.1) as
$$
Y=\sin(2X)+\varepsilon,
$$
where $X \sim U[-3,3]$ is one dimensional covariate and the error $\varepsilon$ has a $N(0,0.2^2)$ law. The Cubic Spline utilizes the third-degree polynomial to fit the regression function piecewise. The regression function can be represented as a linear combination of the truncated power basis,  $r(x)=\gamma^\top B(x)$, where $B(x)=(1, x, x^2, x^3,(x-t_1)_+^3, \cdots,(x-t_{kn})_+^3)^\top$, $a_+=\max\{0, a\}$ and $(t_1, \cdots, t_{kn})$ are knots in some interval $(a,b)$. Thus, the spline estimate of $r(x)$ can be written as
$$
\widehat{r}(x)=\widehat{\gamma}^\top B(x),
$$
where $\widehat{\gamma}=\arg\min_\gamma\sum_{i=1}^n\left(Y_i-\gamma^\top B(X_i)\right)^2$. With the estimating function $U\left(\gamma; y, x\right)=y-\gamma^\top B(x)$, our RWS estimate is
$$
\widehat{r}_k(x)=\widehat{\gamma}_k^\top B(x),\ \widehat{\gamma}_k=\left(\mathbf{B}_{k-1}+\sum_{i\in {\bm i}_k}B(X_i)B(X_i)^\top\right)^{-1}\left(\mathbf{V}_{k-1}+\sum_{i\in {\bm i}_k}B(X_i)Y_i\right),
$$
where $\mathbf{B}_{k-1}=\sum_{j=1}^{k-1}\sum_{i\in {\bm i}_j}B(X_i)B(X_i)^\top$ and $\mathbf{V}_{k-1}=\sum_{j=1}^{k-1}\sum_{i\in {\bm i}_j}B(X_i)Y_i$. Formally, $\widehat{\gamma}_k$ can be expressed as $$
\widehat{\gamma}_k=\left(\sum_{j=1}^{k}\sum_{i\in {\bm i}_j}B(X_i)B(X_i)^\top\right)^{-1}\left(\sum_{j=1}^{k}\sum_{i\in {\bm i}_j}B(X_i)Y_i\right).
$$
This implies that  the RWS estimator is equal to the Cubic Spline estimator provided the same knots.
In this simulation, we compare the following estimators:
\begin{itemize}
\item[1)] Our online updating estimator with full data knot number $kn_f$, denoted by RWS$_{kn_f}$.
\item[2)] Our online updating estimator with the first data batch knot number $kn_1$, denoted by RWS$_{kn_1}$.
\item[3)] The full data Cubic Spline estimator with full data knot number $kn_f$, denoted by CSP$_{f}$.
\item[4)] The simple average of each batch Cubic Spline estimator, denoted by CSP$_{a}$.
\end{itemize}
In the above, we use equidistant knots. The full data knot number $kn_f$ is chosen by Cross-Validation criterion from the full dataset, and the first data batch knot number $kn_1$ is chosen by Cross-Validation criterion from the first data batch. 
The simulation results are reported in Table \ref{tab:7} and Table \ref{tab:8}. We have the following findings:
\begin{itemize}
\item [1)] Our RWS$_{kn_k}$ is much better than CSP$_a$ in the sense that the MISE of RWS$_{kn_k}$ is significantly smaller than that of CSP$_a$; and moreover, our RWS$_{kn_k}$ behaves in the same way as that of the CSP$_f$ when the sample size $n$ is large enough.
\item [2)] Our RWS$_{kn_f}$ and RWS$_{kn_k}$ are robust to the varying of batch size $|{\bm i}|_j$, as the sample size is fixed as $n=12000$.
\item [3)] Our RWS$_{kn_f}$ and CSP$_f$ have the same performance.
\end{itemize}

\begin{table}[htbp]
\centering
\caption{\small The MISE of different estimators under model 2 are summarized over 200 replications, with fixed $n=12000$ and varying batch sizes $|{\bm i}|_j$.
}
\label{tab:7}
\begin{tabular}{c|c|c|c|c}
\hline
$|{\bm i}|_j$ & CSP$_f$ & CSP$_a$ & RWS$_{kn_f}$ & RWS$_{kn_k}$ \\
\hline
30   & $1.856\times 10^{-5}$ & $1.580\times 10^{-3}$ & $1.856\times 10^{-5}$ & $1.995\times 10^{-5}$  \\
\hline
50   & $1.856\times 10^{-5}$ & $1.783\times 10^{-4}$ & $1.856\times 10^{-5}$ & $1.995\times 10^{-4}$ \\
\hline
100  & $1.856\times 10^{-5}$ & $4.012\times 10^{-5}$ & $1.856\times 10^{-5}$ & $1.995\times 10^{-5}$ \\
\hline
300  & $1.856\times 10^{-5}$ & $2.158\times 10^{-5}$ & $1.856\times 10^{-5}$ & $1.995\times 10^{-5}$ \\
\hline
500  & $1.856\times 10^{-5}$ & $2.033\times 10^{-5}$ & $1.856\times 10^{-5}$ & $1.995\times 10^{-5}$ \\
\hline
1000 & $1.856\times 10^{-5}$ & $2.015\times 10^{-5}$ & $1.856\times 10^{-5}$ & $1.995\times 10^{-5}$ \\
\hline
\end{tabular}
\end{table}

\begin{table}[htbp]
\centering
\caption{\small The MISE of different estimators under model 2 are summarized over 200 replications, with fixed batch size $|{\bm i}|_j=100$ and varying $n$.
}
\label{tab:8}
\begin{tabular}{c|c|c|c|c}
\hline
$n$ & CSP$_f$ & CSP$_a$ & RWS$_{kn_f}$ & RWS$_{kn_k}$ \\
\hline
1000   & $1.754\times 10^{-4}$ & $2.663\times 10^{-4}$ & $1.754\times 10^{-4}$ & $2.334\times 10^{-4}$ \\
\hline
10000  & $2.136\times 10^{-5}$ & $4.204\times 10^{-5}$ & $2.136\times 10^{-5}$ & $2.539\times 10^{-5}$ \\
\hline
100000 & $2.493\times 10^{-6}$ & $1.866\times 10^{-5}$ & $2.493\times 10^{-6}$ & $2.733\times 10^{-6}$ \\
\hline
1000000& $3.232\times 10^{-7}$ & $1.714\times 10^{-5}$ & $3.232\times 10^{-7}$ & $3.233\times 10^{-7}$ \\
\hline
\end{tabular}
\end{table}